# A copper sulfide-hydroxypropyl β-Cyclodextrin-reduced graphene oxide composite for highly sensitive electrochemical detection of 5-hydroxytryptamine in biological samples

Aravindan Santhan[ab], Kuo Yuan Hwa*[ab], Slava V Rotkin[c], Cheng-Han Wang[b], Chun-Wei Ou[b]

[a]Department of Molecular Science and Engineering, National Taipei University of Technology, Taipei, Taiwan (ROC)

[b]Graduate Institute of Organic and Polymeric Materials, National Taipei University of Technology, Taipei, Taiwan (ROC).

[c]Materials Research Institute and Department of Engineering Science & Mechanics, The Pennsylvania State University, Pennsylvania 16802, United States.

***Corresponding author**

Professor. Kuo-Yuan Hwa, Email: kyhwa@mail.ntut.edu.tw

Phone number: 02-27712171 ext.2419 (0), 2439, 2442 (lab).

**Abstract**

The precise identification of neurotransmitters is essential for comprehending cerebral function, detecting neurological conditions, and formulating successful therapeutic approaches. The present work investigates the electrochemical detection of serotonin with the excellent hybrid electrocatalyst $Cu_2S$/Hβcd-rGO. $Cu_2S$, with its significant features as improved catalytic activity and enhanced charge transfer when combined with Hβcd-rGO, will enhance the performance. The integration of $Cu_2S$ with Hβcd-rGO, regulated by the van der Waals force and the electrostatic interaction, makes it a stable catalyst without disrupting the composite structure. Also, the aggregation of the $Cu_2S$/Hβcd with the layered sheets of rGO can be highly reduced and resulting in the improvement of the conductivity. Thus, the above features resulted in the improved oxidation response current when fabricated over the glassy carbon electrode (GCE). The SR showed sensitive response at a broad linear range of 0.019 to 0.299 µM and 4.28 to 403.14 µM, resulting in a lower limit of detection (LOD) of 1.2 nM or 0.0012 µM and a sensitivity of about 15.9 $\mu A\ \mu M^{-1}\ cm^{-2}$. The sensor demonstrated excellent selectivity against common interferents, including aminophenol, dopamine, epinephrine, hydroquinone, melatonin, and chlorine. The real sample studies in the biological samples show good recovery values, showing the effectiveness of the as-fabricated sensor. Thus, the cost-efficient and straightforward integration of $Cu_2S$/Hβcd-rGO will be an outstanding electrocatalyst for detecting SR.

**Keywords:** Neurotransmitter; Biological samples; Copper sulfides; hydroxypropyl β-Cyclodextrin; Modified graphene.

## Graphical abstract

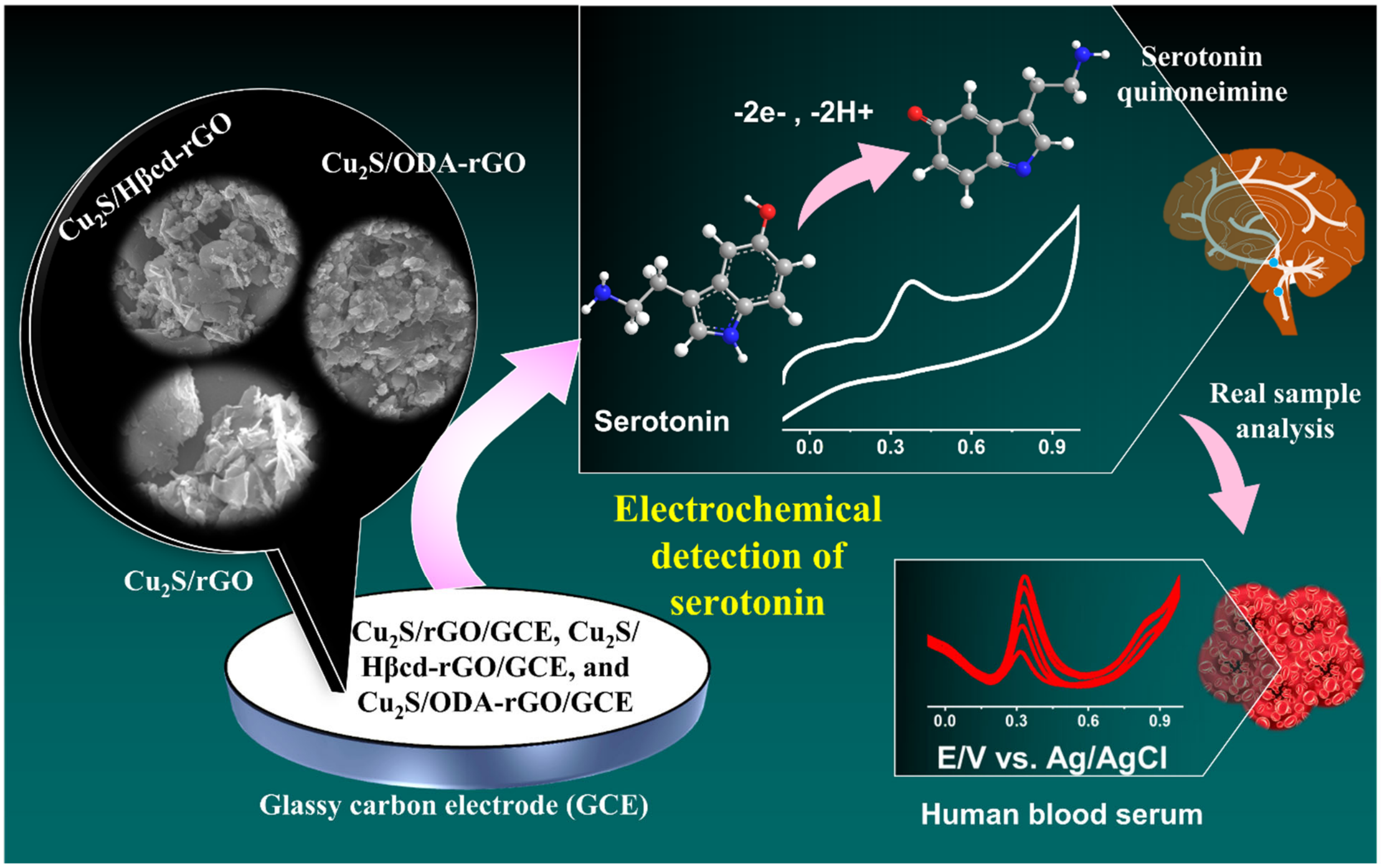

## 1. Introduction

The choice of components for the working electrode is crucial in electrochemical sensors to improve the oxidation process of an analyte, which enhances the electrochemical performance. The 2D heterogeneous materials are a huge focus in electrochemistry due to the atomic scale thickness, unique chemical and electronic properties, as well as higher surface area. The high surface-to-volume ratio, with a highly tunable nature, for improving the performance of the desired study, and higher electron transfer, grabs attention in sensor technology. Such materials include graphene and its reduced graphene oxide, transition metal sulfides, boron nitride, and many more. When taking the transition metal sulfides, including copper, nickel, silver, vanadium, molybdenum, tungsten, and iron, have gained significant interest in electrochemical sensor technologies due to their good electrical conductivity, catalytic activity, and tunable electronic properties as 2D material [1,2], making them suitable for detecting various analytes, including biomolecules, neurotransmitters, pharmaceutical drugs, environmental pollutants, and heavy metal ions. Their diverse morphologies and high surface-to-volume ratios enhance sensitivity, selectivity, and stability in electrochemical sensing [3–10].

Copper sulfide ($Cu_2S$) is a p-type semiconductor that garnered considerable interest owing to an adjustable bandgap, good electrical conductivity, and catalytic activity attractive for electrochemical sensors [10–14]. A recent study by Kuo-Yuan Hwa *et al.,* reported the significance of utilizing $Cu_2S$ for the electrochemical sensing application. The $Cu_2S$ holds numerous properties that uphold enhanced and stable response towards the analytes exposed for detection. Several composite materials that integrate copper sulfide and carbon-derived nanomaterials, including graphene derivatives, carbon black, carbon nanotubes, activated carbon, carbon nanofibers, and coal, have been examined for sensors [15,16]. In this work, oxidized graphene oxide (GO) was reduced through chemical or physical treatment, yielding reduced graphene oxide (rGO) materials [17]. rGO is a common 2D material with strong interaction with adsorbed molecules that can result in significant charge transfer. Potential re-aggregation of rGO might happen post-reduction due to van der Waals interactions, leading to insufficient solubility of rGO and impeding sensor fabrication. Chemical modification of rGO is used to mitigate the solubility issues [18,19].

Cyclodextrins are oligosaccharides consisting of six to eight glucose units connected by 1,4-glucosidic bonds. β-cyclodextrin (βcd) exhibits notable chemical-based selectivity for specific molecules. It establishes enduring host-guest complexes with substantial binding affinity above $10^3$. The oligosaccharide molecules possess a hydrophobic core and a hydrophilic exterior rich in hydroxyl groups. βcd dissolves in water and can be used to improve the solubility and stability of rGO in solution. Hydroxypropyl-beta-cyclodextrin (Hβcd) is a derivative of βcd, altered by the addition of hydroxypropyl groups to improve solubility in water and other solvents [20–23].

Octadecyl amine (ODA), an extended-chain fatty acid alkyl amine noted is known to engage with rGO without necessitating an additional binding agent, act as a stabilizer, and provide sufficient dispersion in organic compounds and non-polar solutions [24–26]. Among the above integration of the material to improve the efficiency, stability, and good adsorption of the analyte is possible due to the vandewaals force that exists with other bonding. The rate of aggregation when integration of $Cu_2S$/Hβcd-rGO is highly reduced and retains the structure of the composite without any ruptures. The synergistic response resulted from the combination of different materials with unique features, wherein the $Cu_2S$/Hβcd-rGO will enhance the sensitivity, the oxidation response of SR, and lower the overpotential. Thus, the integration of $Cu_2S$ with rGO significantly augments its electrochemical efficacy by enhancing electron exchange kinetics and offering an extensive surface area for analyte interaction. Moreover, the $Cu_2S$/Hβcd-rGO will be an active catalyst that boosts the electrode's performance with excellent outcomes.

It is well known that neurotransmitters play a significant role in controlling essential biological and physiological functions of the human body [27,28]. Serotonin (SR), a vital monoamine neurotransmitter, serves as one of the catecholamines that, arguably, has received the most attention recently [29,30]. Serotonin (5-hydroxytryptamine) significantly influences various physiological processes, such as regulating mood, hunger management, patterns of sleep, cardiovascular functioning, and the neuroendocrine signals [21,31,32]. Biological fluids exhibit varying concentrations of serotonin [33,34]. The central nervous system, the digestive system, and the blood platelets are the primary locations where it is observed. For example, the level of serotonin in human blood may vary from 101 to 283 ng $m^{-1}$ $L^{-1}$, while in serum it can range from 0.57 to 2 μM, in urine it can range from 0.295 to 0.687 μM, and in cerebral fluid it spans from 0.8 to 3.7 nmol $L^{-1}$. There is a correlation between abnormal serotonin concentrations and a variety of medical disorders [35,36]. Insufficient level of serotonin was linked to conditions such as depression, as well as to serious diseases such as Parkinson's disease, Alzheimer's disease, and migraines. At the same time, an elevated level of serotonin was linked to conditions such as serotonin syndrome, also called irritable bowel syndrome, and fevers that are difficult to regulate. Thus, the determination of neurotransmitters in the biofluids is an important diagnostic tool, allowing the detection of several disorders. This requires methods of detection that are stable, accurate, and relatively quick to perform [37,38]. The electrochemical techniques were utilized to identify SR in the past, with the electrocatalyst materials performing a critical role in this detection. A variety of electrocatalysts was proposed to match various target analytes. Still, developing cost-effective electrode materials that demonstrate superior electrical conductivity, enduring stability, and unique functionalities is a challenge [39–41]. Our group has previously studied needle-like $Cu_2S$ structures for the efficient detection of serotonin with a limit of detection of about 3.2 nM for a linear range of 0.029 to 607.6 μM [10]. Also, our study on $Zn_2P_2O_7$/NbC modified over GCE for the detection of serotonin in biological samples exhibited an LOD of 0.0055 μM for a wide linear range of 0.019 to 563.68 μM [42]. Even though SR detection has been studied, in order to improve the LOD and to improve the electrochemical performances, we have incorporated rGO, Hβcd-rGO, and ODA-rGO with $Cu_2S$ rather than studying $Cu_2S$ individually as reported before. Accordingly, $Cu_2S$/Hβcd-rGO electrocatalyst showed excellent features that provide outstanding electrochemical responses, as evidenced and given in the present study. Several other reports have been established with the utilization of 2D materials for SR detection. A report on $FeVO_4$ nanoflakes decorated on $Ti_3C_2$ Mxene as a nanocomposite for the detection of SR with LOD about 5.88 nM for the linear range of 25 to 750 nM [43]. Another report on the electrochemical sensing of SR with exfoliated graphite/polythiocyanogen composite resulted in an LOD of 59.5 nM [44]. A study on SR with graphitic carbon nitride-$MoS_2$ with 3D graphene as the composite material showed 15 nM of LOD for a linear range of 1.8–3.8: 53.8–3693.8 μM [45]. But, the present work with a lower LOD is employed as an excellent electrocatalyst material for SR detection in comparison with other 2D-based composite electrode materials.

Moreover, combining $Cu_2S$ with the reduced graphene oxide (rGO), hydroxypropyl beta-cyclodextrin (Hβcd), and octadecylamine (ODA) functionalized rGO, provides enhanced selection, stability, and sensitivities for the detection of neurotransmitters, as will be shown next with the electrochemical sensing of SR.

## 2. Materials and Methodology

Details of the preparation of composite materials to be placed on a glassy carbon electrode (GCE) for the sensing of SR are described in the Supporting Information: **S1.** materials and reagents, **S2.** preparation of copper sulfide, **S3.** reduced graphene oxide, **S4.** octadecyl amine functionalized rGO, **S5.** hydroxypropyl-beta-cyclodextrin modified rGO. **S6.** presents instrumental information, **S7.** morphological analysis for Cu2S and modified rGO. Overall synthesis scheme illustrated in **Scheme. 1.**

**2.1. Preparation of $Cu_2S$/rGO:**

200 mg copper sulfide ($Cu_2S$) and 100 mg of rGO were added into a 75 ml solution (50 ml DW and 25 ml ethanol), and the solution was subjected to an ultrasonication process for 1 h to disperse the materials and later allowed to stir for 12 h. Then, the obtained product is centrifuged with DW and ethanol at 5000 Rpm for 15 min several times to eliminate the unreacted particles. The resultant sample was allowed to undergo the calcination process in a hot air oven at 60 ºC.

**2.2. Preparation of $Cu_2S$/ODA-rGO:**

200 mg of copper sulfide ($Cu_2S$) and 100 mg of ODA-rGO were taken together and added into the 75 ml solution (50 ml DW, 25 ml ethanol), for an ultrasonication process for around 1 h to disperse and mix, and later allowed to stir for 12 h. The mixture solution was centrifuged several times with water and ethanol to remove the impurities. The resultant sample was dried in a hot air oven at 60 ºC.

**2.3. Preparation of $Cu_2S$/Hβcd-rGO;**

200 mg of copper sulfide ($Cu_2S$) and 100mg of Hβcd-rGO were added together in a 75 ml solution (50 ml DW, 25 ml ethanol) for ultrasonication for 1 h to disperse the materials; later allowed for magnetic stirring for 12 h. The colloidal solution was centrifuged many times with water and ethanol to fully eliminate the impurity materials. The resultant sample was dried in a hot oven at 60 ºC.

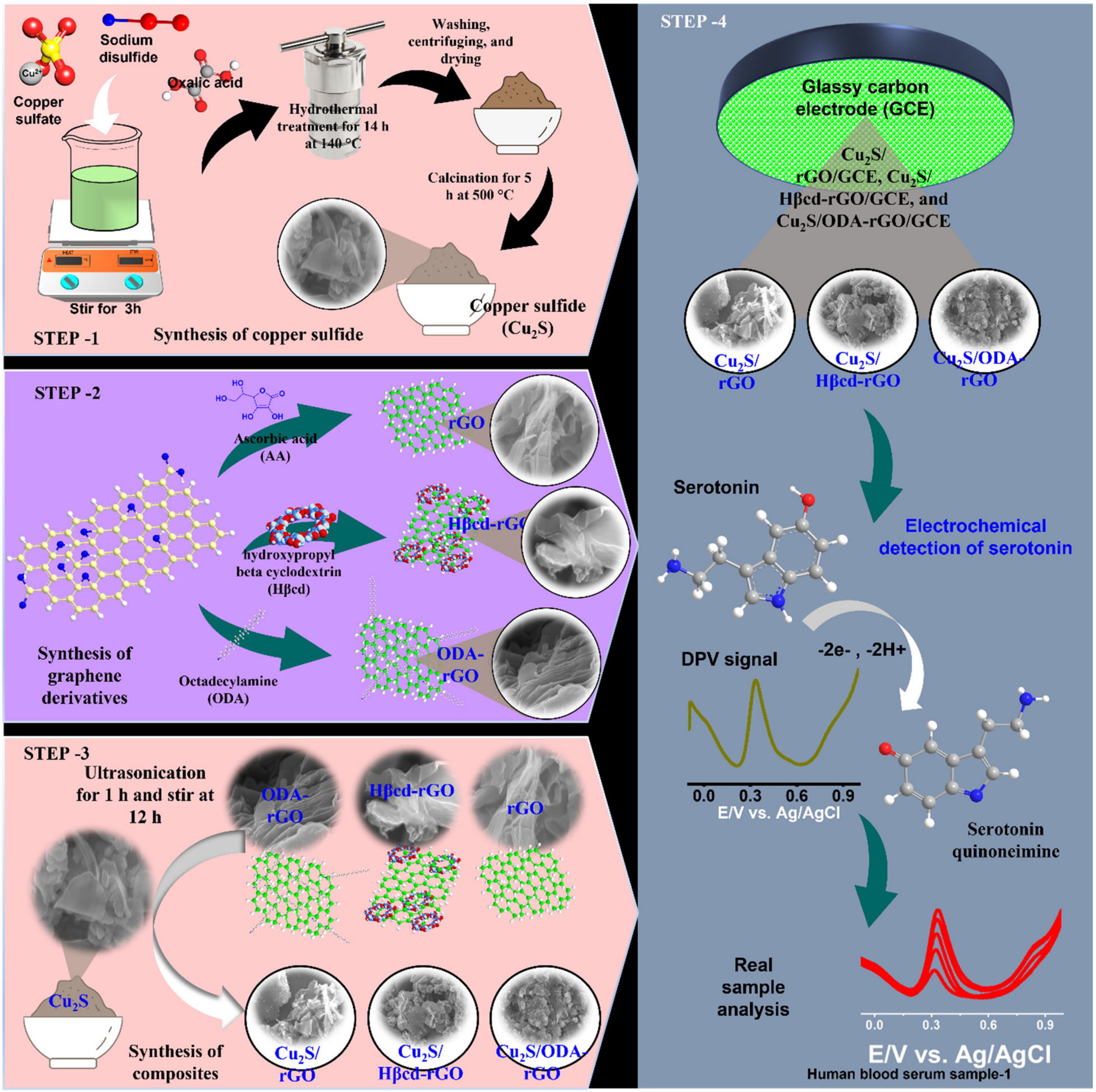

**Scheme.1** Synthesis scheme of $Cu_2S$, rGO, Hβcd-rGO, ODA-rGO, $Cu_2S$/rGO, $Cu_2S$/Hβcd-rGO, $Cu_2S$/ODA-rGO and its electrochemical applications.

### 2.6. Fabrication of GCE:

The glassy carbon electrode (GCE) was used as a working electrode. The surface of the electrode must be cleaned before fabricating the material onto the GCE. The surface of GCE was polished with alumina slurry and cleaned with water and ethanol several times to eliminate the impurity particles. After that, 3 mg/mL (6 µL) of synthesized $Cu_2S$, rGO, Hβcd-rGO, ODA-rGO, $Cu_2S$/rGO, $Cu_2S$/Hβcd-rGO, and $Cu_2S$/ ODA-rGO materials were drop-coated over the bare GCE and allowed to dry in a hot air oven for 10 min.

## 3. Results and discussion

### 3.1. XRD

The X-ray diffraction analysis was used to determine the crystalline structure and lattice parameters of the synthesized materials. The **Figure. 1(A)** illustrates the XRD pattern of $Cu_2S$ at a 2θ range of 15° to 70°. The crystal structure was confirmed with the (JCPDS NO: 01-072-1071) for a tetragonal crystal system with space group of P43212 with the lattice unit cell parameters: a = 3.9960 Å, b = 3.9959 Å, c = 11.2869 Å, α=β=γ values are 90°. The following *hkl* planes for $Cu_2S$ (101), (102), (110), (111), (112), (104), (113), (200), (201), (202), (203), (212), (204), (220) and (215) correspond to two theta values: 23.5°, 27.6°, 31.6°, 32.1°, 34.9°, 39.0°, 39.7°, 45.3°, 45.9°, 48.1°, 51.6°, 53.6°, 54.4°, 66.5°, and 67.1° respectively. The obtained XRD results confirm the copper sulfide with no other impurities. **Figure. 1(B)** illustrates the XRD pattern of rGO, which was further functionalized with Hβcd and ODA materials. The XRD pattern of rGO is confirmed based on the two peaks, one obtained at 25.4°, corresponding to the (002) plane [46,47]. The XRD pattern of Hβcd-rGO was confirmed based on peaks present at 25.8° for the (002) plane, typically present after the reduction of graphene oxide (GO) and removal of oxygen-functionalized groups. The peaks present at 42.6° is attributed to the rGO peak associated with in-plane structure of graphene sheets. The XRD pattern of ODA-rGO displays a characteristic diffraction peak at 22.3° of (002) plane and a low intensity peak noticed at 42.7º [24,25,48]. This may be related to the aggregation of ODA-rGO sheets together. **Figure. 1(C)** depicts the XRD pattern of $Cu_2S$/rGO, $Cu_2S$/Hβcd-rGO, and $Cu_2S$/ODA-rGO composite materials that display peaks similar to and obtained from the original $Cu_2S$ and rGO material, confirming the formation of $Cu_2S$/rGO composites. Some broadening of the peaks might be due to increased stacking disorder of graphene layers and the defects introduced in $Cu_2S$ upon binding with functionalized rGO. Which may lead to better electron transfer via improved active surface sites/area. The average crystalline size of the synthesised materials was calculated using Scherrer's formula ($D = k\lambda/\beta\cos\theta$) where k, λ, β, and θ correspond to the Scherrer constant (related to the shape of the particle), the X-ray wavelength (we used Cu/Kα= 0.154 nm), full-width half maxima of the high intensity plane, and Bragg's diffraction angle, respectively. The calculated average crystalline size for the synthesized $Cu_2S$, rGO, Hβcd-rGO, ODA-rGO $Cu_2S$/rGO, $Cu_2S$/Hβcd-rGO, and $Cu_2S$/ODA-rGO composites was 68.4 nm, 13.0 nm, 8.2 nm, 11.8 nm, 43.9 nm, 32.4 nm, and 38.2 nm, respectively. The reduced crystalline size of the sample will provide a higher density of surface atoms and defect sites, which act as active centers for electrochemical reactions, thereby enhancing sensitivity and lowering the detection limit.

We estimated the dislocation density (δ) and lattice strain (ε) from XRD data using the formula:$\delta = \frac{1}{D2}$ and $\varepsilon = \beta/4\tan\theta$. The calculated dislocation densities were δ= 0.21 $m^{-2}$, 5.36 $m^{-2}$, 14.51 $m^{-2}$, 7.16 $m^{-2}$, 1.77 $m^{-2}$, 1.32 $m^{-2}$, and 0.93 $m^{-2}$, and the lattice strain ε = 1.59, 11.54, 18.70, 15.16, 4.46, 3.70, and 3.53 for $Cu_2S$, rGO, Hβcd-rGO, ODA-rGO $Cu_2S$/rGO, $Cu_2S$/Hβcd-rGO, and $Cu_2S$/ODA-rGO composites, respectively. Such dislocation density and the lattice strain will offer additional electrochemically active sites and reduce the energy barrier charge-transfer resistance.

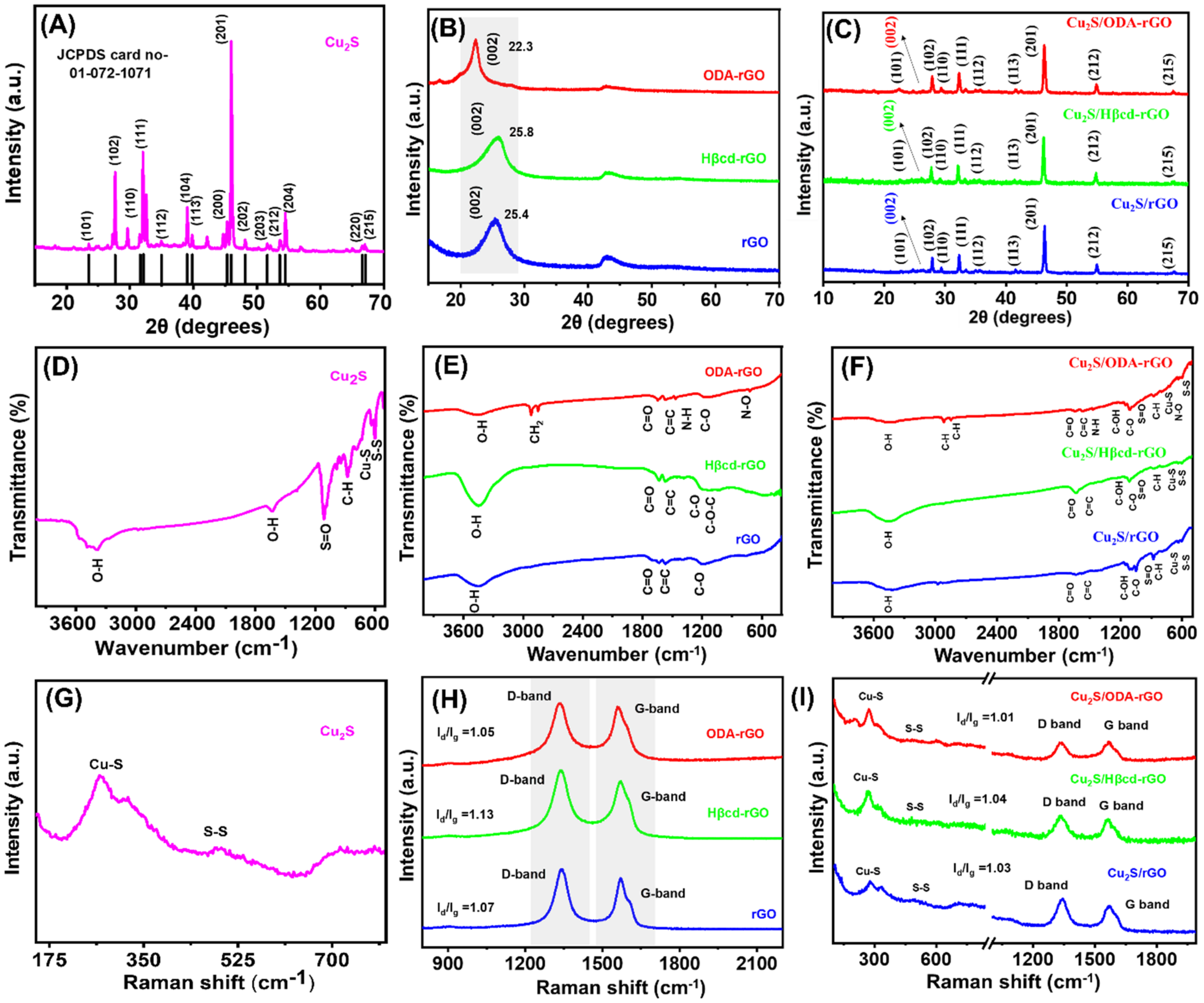

**Figure. 1(A)** XRD spectra of Cu2S, **(B)** XRD spectra of rGO, Hβcd-rGO, and ODA-rGO, **(C)** XRD spectra of Cu2S/rGO, Cu2S/Hβcd-rGO, and Cu2S/ODA-rGO, **(D)** FTIR image of Cu2S, **(E)** FTIR images of rGO, Hβcd-rGO, and ODA-rGO, **(F)** FTIR images of Cu2S/rGO, Cu2S/Hβcd-rGO, and Cu2S/ODA-rGO, **(G)** Raman image of Cu2S, **(H)** Raman images of rGO, Hβcd-rGO, and ODA-rGO, **(I)** Raman images of Cu2S/rGO, Cu2S/Hβcd-rGO, and Cu2S/ODA-rGO.

### 3.2. FT-IR spectrum:

The Fourier transform infrared spectroscopy (FT-IR) analysis was used to determine the functional groups and chemical bonds present in the synthesis material. **Figure. 1(D)** shows the FT-IR spectrum for $Cu_2S$ material at a frequency range of 400 to 4000 $cm^{-1}$. Three characteristic vibration peaks around 520, 608, and 1065 $cm^{-1}$ belong to $Cu_2S$, attributed to S-S bisulfide group (520 $cm^{-1}$), Cu-S stretching modes (608 $cm^{-1}$), and an asymmetric valence S=O stretching vibration (1065 $cm^{-1}$). The peak present at 867 $cm^{-1}$ was related to the aromatic bending vibration of C-H. The peaks at 1670 and 2365 $cm^{-1}$ correspond to bending mode of O-H group of a water molecule [49]. **Figure. 1(E)** presents the FT-IR spectrum for rGO and functionalized Hβcd-rGO and ODA-rGO materials. The rGO peak at 1020 $cm^{-1}$ is attributed to the alkoxy C-O stretching vibrations, the peak at 1716 $cm^{-1}$ corresponds to an aromatic stretching vibration of C=O group. The peak at 1558 $cm^{-1}$ belongs to stretching vibration of C=C group and the broad -peak near 3437 $cm^{-1}$ is due to the O-H stretching vibration [50]. The Hβcd-rGO materials show both the characteristic peaks of Hβcd and rGO, the O-H stretching vibration (3437 $cm^{-1}$), the vibrations of saccharide structure from βcd: the antisymmetric glycosidic C-O-C stretching (1155 $cm^{-1}$) and the coupled C-O stretching vibration (1030 $cm^{-1}$), the rGO vibrations: C=C stretching vibration (1556 $cm^{-1}$), C=O (1717 $cm^{-1}$) [20,21].

The FT-IR spectrum of ODA-rGO, **Figure. 1(E)**, in addition to the features corresponding to both rGO and ODA groups, exhibited the peaks at 783 $cm^{-1}$ and 1434 $cm^{-1}$ related to N-O group and N-H bonds (in addition to carboxylic and hydroxyl groups). The asymmetric peaks at 2845 $cm^{-1}$ are due to the alkyl group stretching vibrations of $CH_2$ bonds in methyl and methylene groups. The peaks at 3437, 1020, and 1558 $cm^{-1}$ belong to rGO stretching vibrations of O-H bond, alkoxy C-O stretching vibrations, and C=C bond stretching respectively. The peak at 1716 $cm^{-1}$ is due to the C=O amide bond. Thus, the obtained peaks confirmed the formation of ODA-rGO materials [24,51].

**Figure. 1(F)**, shows the $Cu_2S$/ODA-rGO composite material displays peaks similar to those obtained from the bare $Cu_2S$ and ODA-rGO, though the formation of the $CH_2$ methylation

group of functionalized ODA, split into a doublet for the stretching and deformation vibration, is noted. The FT-IR spectrum for $Cu_2S$/Hβcd-rGO shows mostly peaks similar to those in bare $Cu_2S$ and Hβcd-rGO.

**3.3. Raman spectroscopy:**

**Figure. 1(G)** illustrates the Raman spectra for $Cu_2S$ which shows the $A_{1g}$ S-S stretching vibrational mode (484 $cm^{-1}$), and the Cu-S (covellite phase) mode (267 $cm^{-1}$) [52,53]. **Figure. 1(H)** depicts the Raman spectra for rGO and functionalized Hβcd-rGO and ODA-rGO materials. The spectrum shows two characteristic peaks at 1343 $cm^{-1}$ and 1570 $cm^{-1}$ associated with the D and G bands of 2D-carbon. The presence of the D band signifies the existence of defects, functional groups, lattice disorders, and wrinkles in the rGO. Slight G peak shifts (1558 $cm^{-1}$ in ODA-rGO and 1571 $cm^{-1}$ in Hβcd-rGO) and broadening of the G band might be due to the functionalization of rGO material. The D-band of ODA-rGO is at 1331 $cm^{-1}$ compared to 1338 $cm^{-1}$ in Hβcd-rGO.

The $I_D/I_G$ intensity ratio, which is related to the degree of defects and disorders, for rGO, Hβcd-rGO, and ODA-rGO materials is 1.07, 1.05, and 1.13, respectively, possibly resulted from a higher disorder due to the Hβcd functionalization or ODA amines, which are covalently bound to the basal planes of rGO. The higher defect density should facilitate the electron transfer during the electrochemical performance. **Figure. 1(I)** represents the Raman spectrum for $Cu_2S$/rGO, $Cu_2S$/Hβcd-rGO, and $Cu_2S$/ODA-rGO composites, displaying the Raman peaks for both materials. The $I_D/I_G$ ratio calculated for $Cu_2S$/rGO, $Cu_2S$/Hβcd-rGO, and $Cu_2S$/ODA-rGO composites was 1.03, 1.01, and 1.04, respectively. Thus, the Raman-derived augmentation in structural disorder and interfacial coupling directly corroborates the enhanced electrocatalytic performance and will result in enhanced peak current response.

**3.4. UV-spectroscopy analysis:**

The diffused reflectance spectra of $Cu_2S$, rGO, Hβcd-rGO, ODA-rGO, $Cu_2S$/rGO, $Cu_2S$/Hβcd-rGO, and $Cu_2S$/ODA-rGO composites were obtained utilizing a UV-Vis spectrometer in the wavelength range 220–800 nm, as illustrated in **Figure. 2(A)** and **Figure. S1(A, C).** For such microstructured materials, both absorption and scattering must be considered for proper interpretation of reflectance. The diffuse reflectance, R, data is associated with the Kubelka-Munk factor F(R) through the established relationship [54]:

$$F(R) = (1-R)^2/2R$$

At the same time, this factor equals F(R)=Absorption(E)/Scattering(E), within the Kubelka-Munk model (here we assume dipole-allowed optical transitions at the gap and achromatic/energy-independent scattering factor). This allows us to determine the edge of fundamental absorption or the band gap. The band gaps might be affected by several factors, including the environment, pressure, charges, defects, as well as by the structural integrity of the material. The bandgap of $Cu_2S$, rGO, Hβcd-rGO, ODA-rGO, $Cu_2S$/rGO, $Cu_2S$/Hβcd-rGO, and $Cu_2S$/ODA-rGO composite materials is shown in **Figure. 2(B-D)** and **Figure. S1(B, D-F)**. The band gap of bare $Cu_2S$ was found to be 1.22 eV. The rGO material is best fitted to 1.09 eV. In the functionalized materials, the gap was 0.89 eV, 0.95 eV, 0.93 eV, 0.80 eV, and 0.86 eV for Hβcd-rGO, ODA-rGO, $Cu_2S$/rGO, $Cu_2S$/Hβcd-rGO, and $Cu_2S$/ODA-rGO, respectively. Notable Urbach tails were seen in all materials due to a large degree of disorder.

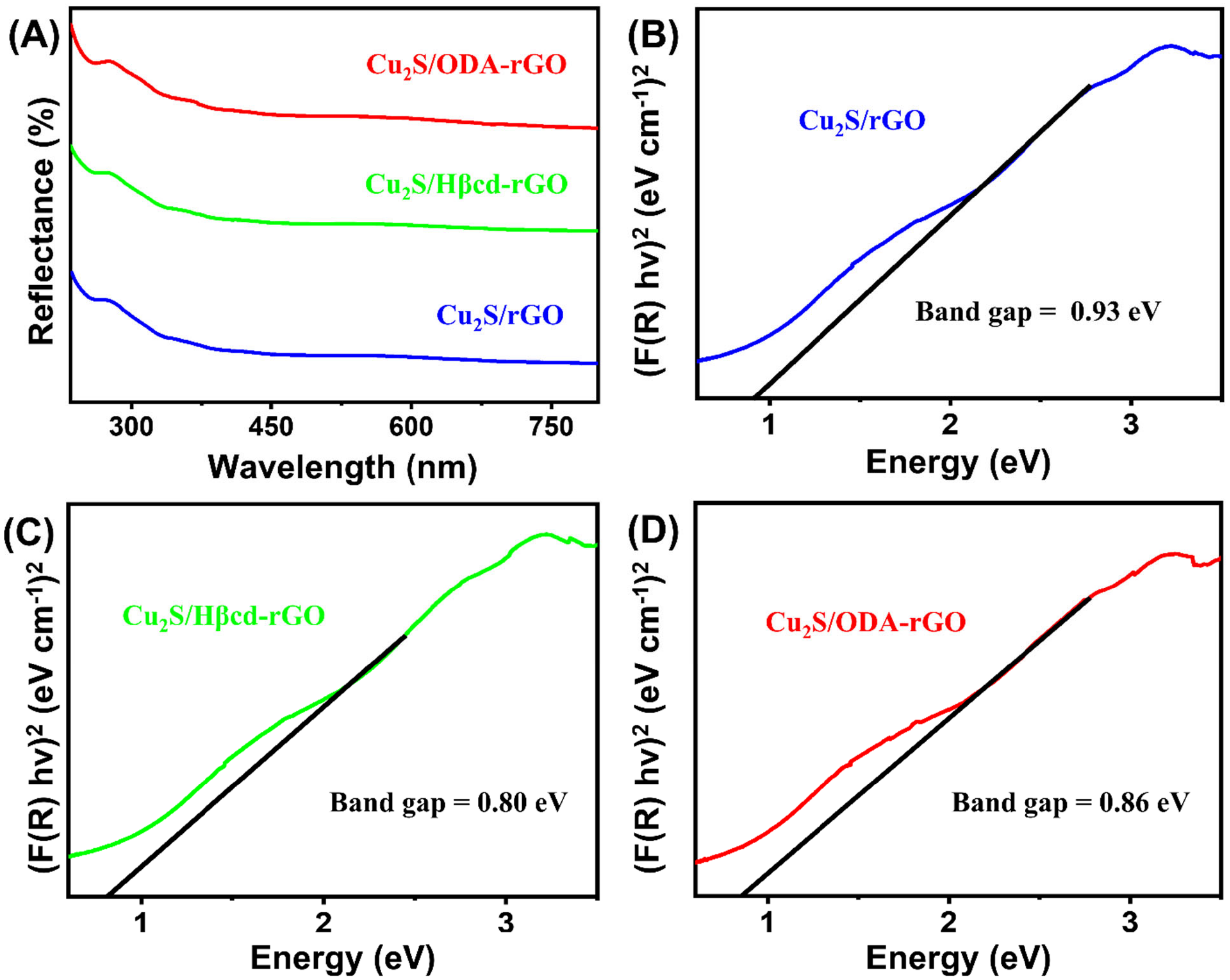

**Figure. 2(A)** UV-Vis's spectra of $Cu_2S$/rGO, $Cu_2S$/Hβcd-rGO, and $Cu_2S$/ODA-rGO, **(B-D)** Kubelka-Munk fitting for bandgap of $Cu_2S$/rGO, $Cu_2S$/Hβcd-rGO, and $Cu_2S$/ODA-rGO.

### 3.5. X-ray photoelectron spectroscopy

The XPS analysis is used to measure the elemental composition and chemical states present in the composite materials. **Figure. 3(A)** illustrates the overall survey of $Cu_2S$/rGO composite and **Figure. 3(B-E)** shows the core level peaks of Cu 2p, S 2p, C 1s, and O 1s. The core level Cu 2p peak was deconvoluted to a spin-orbit split doublet (Cu $2p_{3/2}$ and Cu $2p_{1/2}$) at binding energies 932.5 eV and 952.2 eV. The two satellite peaks at a binding energy of 939 eV and 959 eV are due to the shake-off [10,55]. The core level S 2p peak is also spin-orbit split into S $2p_{3/2}$ and S $2p_{1/2}$ at binding energies of 161.1 eV and 163.0 eV [10,56,57]. The C 1s displays characteristic peaks at 284.4 eV, and 285.9 eV, attributed to the C-C bond, the other peak at 287.3 eV related to the C=C bond, and the 287.3 eV line for the C-O bond. Lower intensity of the latter peak reveals the successful reduction of GO to rGO [50,58,59]. The high-resolution spectra of O 1s were deconvoluted into the peak at 530.2 eV corresponding to Cu-O and the peaks at 532.5 eV and 531.5 eV corresponding to the C-O bond and oxygen vacancies [50,59]. **Figure. 3(F)** depicts the XPS survey spectra for $Cu_2S$/Hβcd-rGO composite and **Figure. 3(G-J)** presents detailed structure for Cu 2p, S 2p, C 1s and O 1s lines. The Cu 2p peaks show slight hybridization (binding energy of 931.5 eV and 951.7 eV), as well as the S 2p peaks (161.2 eV and 162.5 eV). The deconvoluted C 1s core peak presents C-C or C=C bond (283.5 eV), C-O bond (285.7 eV), C=O bond (286.8 eV), and O-C=O bond (287.9 eV). The latter is present in the functional group of Hβcd. The XPS spectra for O 1s consist of three main peaks around 529.0 eV, 530.2 eV, and 531.6 eV, attributed to the metallic oxygen M-O, C=O bond, and O-C=O bond, respectively. **Figure. 3(K)** represents the overall survey spectrum of $Cu_2S$/ODA-rGO composite and the core level peak details for Cu 2p, S 2p, C 1s, N 1s, and O 1s are depicted in **Figure. 3(L-O)**. The Cu 2p (951.8 eV and 931.8 eV), the S 2p (162.5 eV and 161.2 eV) with the satellite peaks (939 eV and 957 eV), and C 1s peaks at 284.2 eV (C-C or C=C), 284.2 eV (amide C-N or hydroxyl groups C-O), 286.7 eV (ether CO-O bond), 288.0 eV (amide CONH-R or carbonyl C=O functional groups) [60,61]. The O 1s peaks at 530 eV (M-O bond), 531.2 eV (C=O bond), and 532.4 eV (O-C=O bond) are similar to other compounds. The XPS spectra of N 1s show a peak at 400 eV, which is ascribed to the amide ($NH_2$) functional group, characteristic of ODA. The other two peaks at 399.5 eV and 402.1 eV correspond to C-N-C and C(O)N bonds, respectively [60,61]. Thus, the surface composition and electronic interactions revealed from XPS elucidate the enhanced electrochemical response, indicating that optimal oxidation states and defect chemistry are fundamental to the high sensitivity and low detection limit of the composite material.

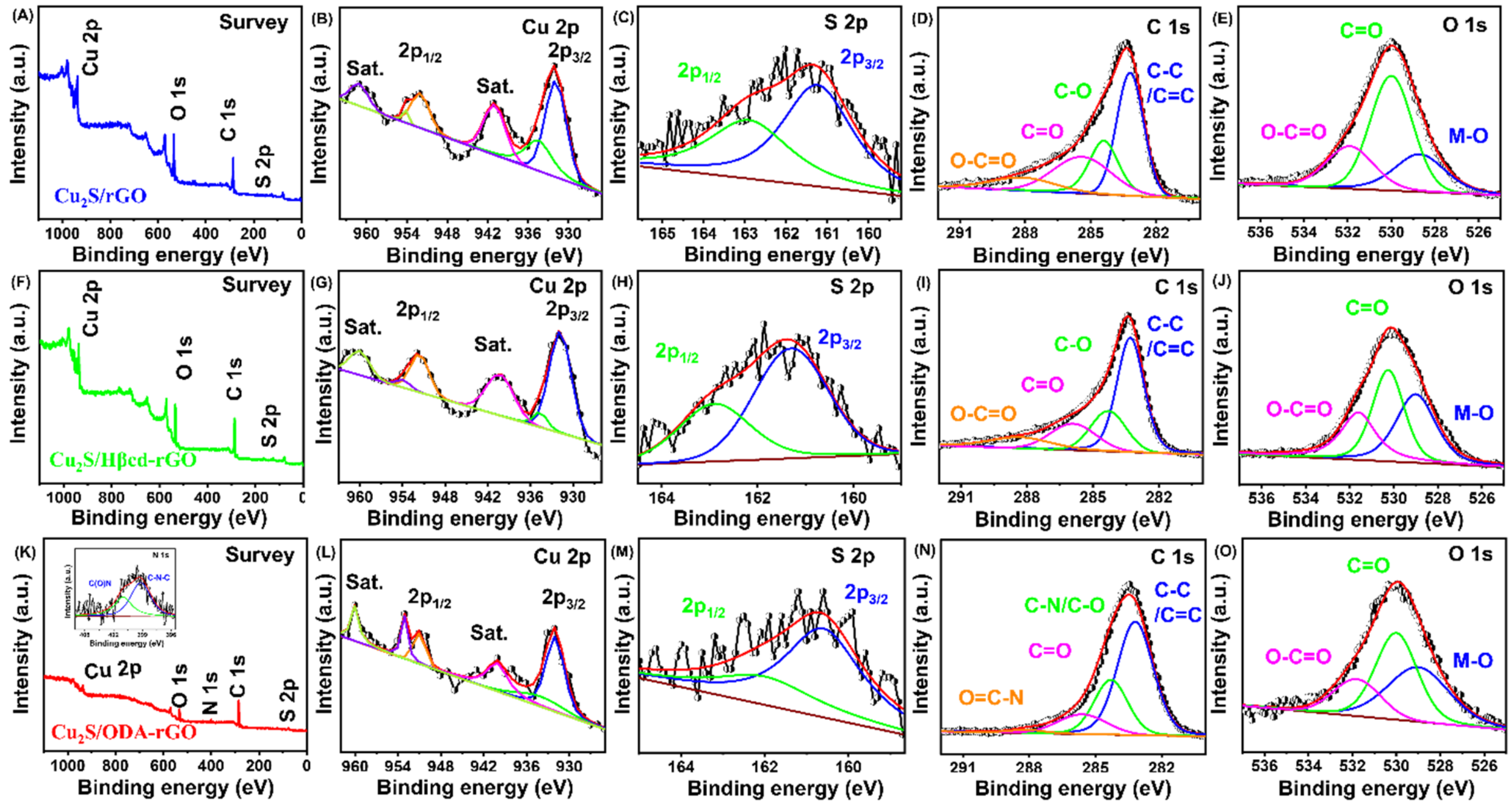

**Figure. 3(A)** XPS overall spectrum of $Cu_2S$/rGO, individual spectrum of **(B)** Cu 2p, **(C)** S 2p, **(D)** C 1s, **(E)** O 1s, **(F)** XPS overall spectrum of $Cu_2S$/Hβcd-rGO, individual spectrum of **(G)** Cu 2p, **(H)** S 2p, **(I)** C 1s, **(J)** O 1s, **(K)** XPS overall spectrum of $Cu_2S$/ODA-rGO, individual spectrum of **(L)** Cu 2p, **(M)** S 2p, **(N)** C 1s, **(O)** O 1s, and **(K)** N 1s(insert).

### 3.6. Morphological analysis

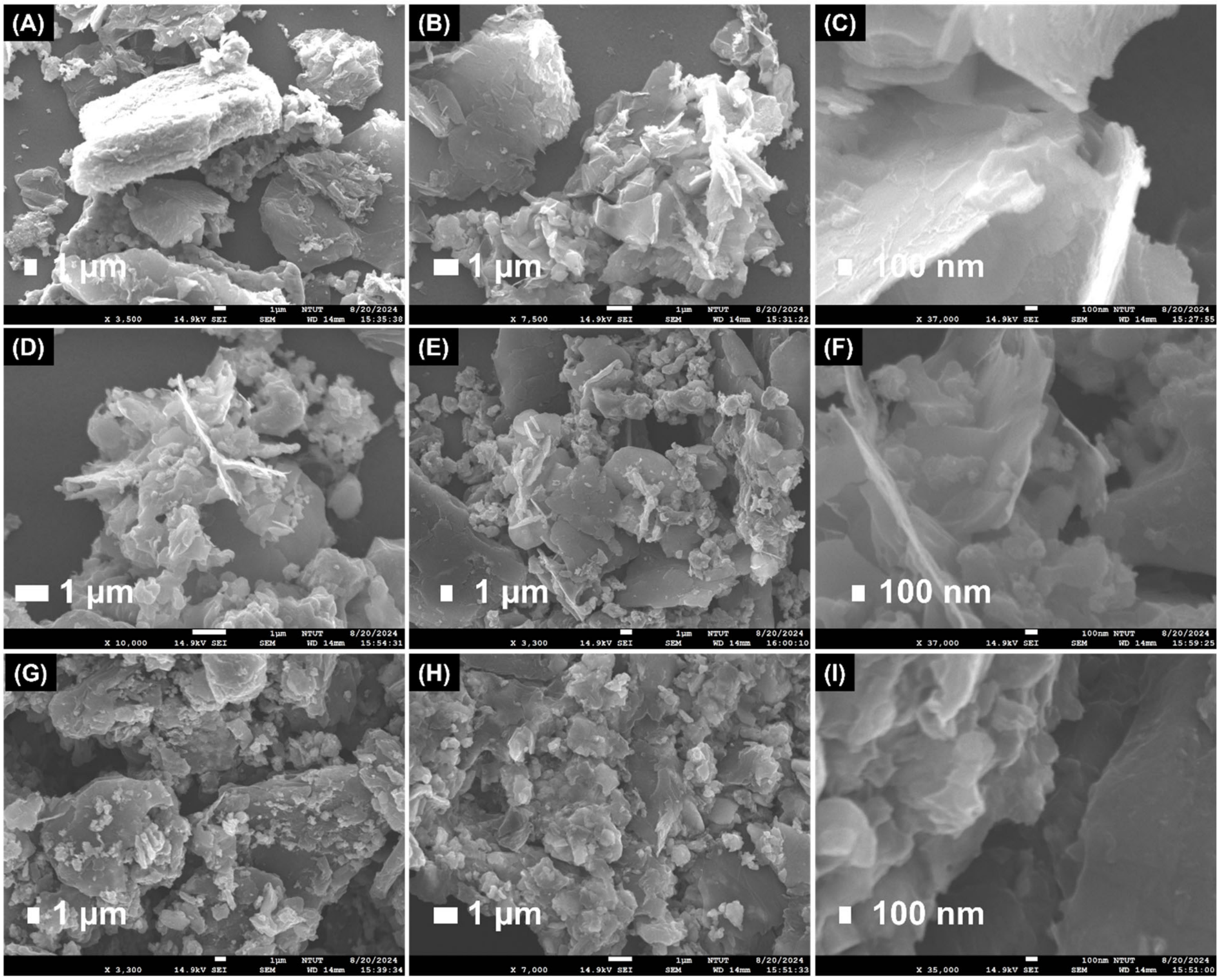

**Figure. 4** Various magnification FESEM pictures of **(A-C)** $Cu_2S$/rGO, **(D-F)** $Cu_2S$/Hβcd-rGO, and **(G-I)** $Cu_2S$/ODA-rGO.

The structural and morphological change of $Cu_2S$/rGO composites is traced here with a set of Field-Emission Scanning Electron Microscopy (FESEM) images of original copper sulfide ($Cu_2S$), original reduced graphene oxide (rGO), hydroxypropyl beta-cyclodextrin-functionalized reduced graphene oxide (Hβcd-rGO), and octadecylamine-modified reduced graphene oxide (ODA-rGO) as given in **Figure 4** and **Figures S2 & S3** (for bare materials). **Figure. 4(A-C)** shows $Cu_2S$ combined with rGO samples, exhibiting a randomly dispersed, irregular morphology characterized by a combination of platelet-like and agglomerate formations. **Figure. 4(C)** illustrates the wrinkled, layered structure of rGO, typical for partially exfoliated graphene sheets. The latter offers an extensive surface area, hence improving charge transfer properties and stability in sensors. **Figure. 4(D-F)** shows $Cu_2S$/Hβcd-rGO composite, exhibiting a more integrated network arrangement. The $Cu_2S$ phase is efficiently distributed, likely owing to improved interaction with cyclodextrin. **Figure. 4(F)** shows the $Cu_2S$ nano-platelets integrated into the rGO matrix, indicating a robust contact between $Cu_2S$ and Hβcd-rGO. **Figure. 4(G-I)** shows $Cu_2S$/ODA-rGO composite material with diminutive, distributed $Cu_2S$ particles (with less agglomeration) incorporated into the rGO layers, indicating enhanced dispersion resulting from ODA modification, which may improve the material stability and enhance the surface interactions.

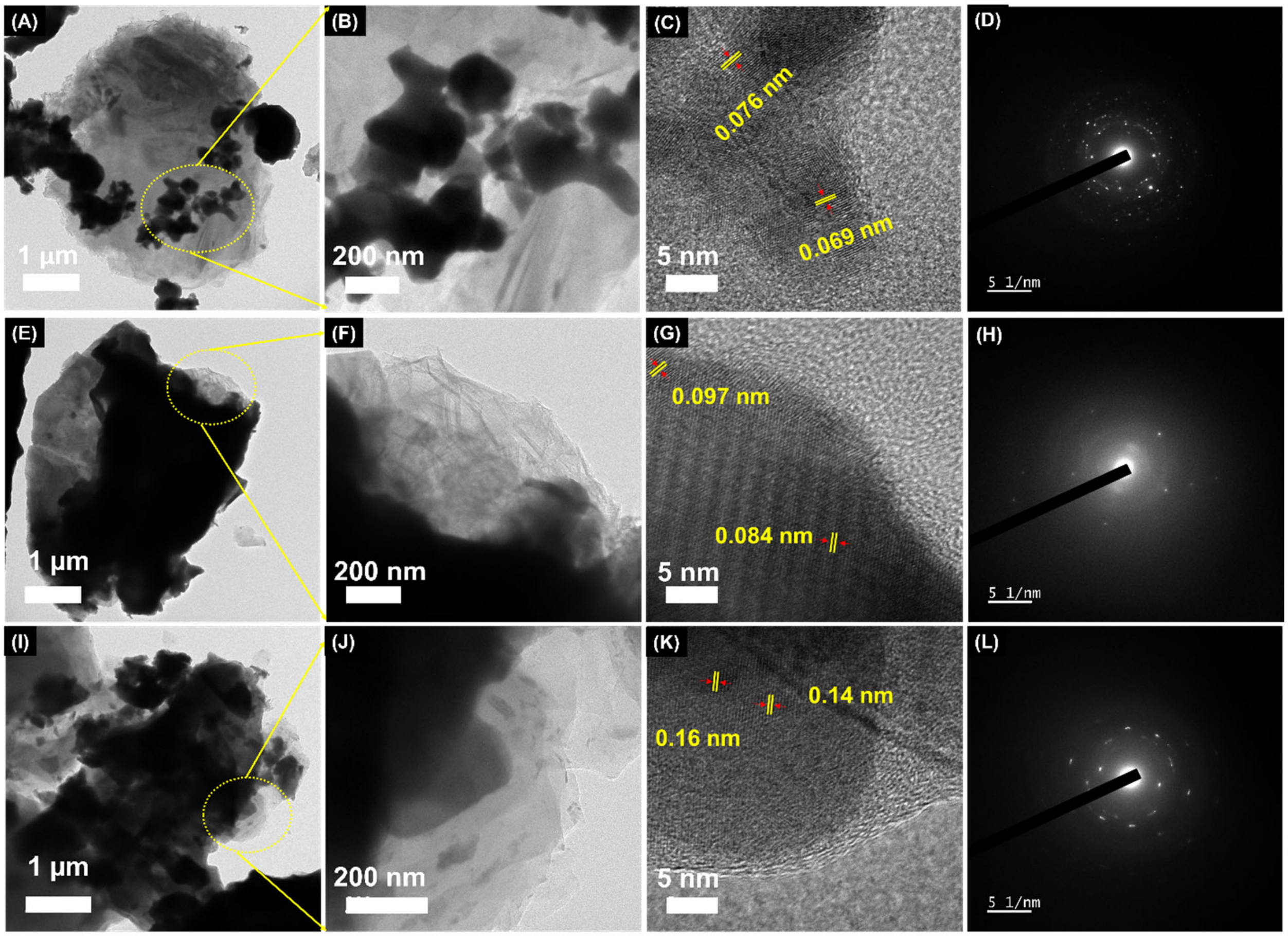

**Figure. 5 (A-C)** $Cu_2S$/rGO TEM pictures and **(D)** SAED pattern images, **(E-G)** $Cu_2S$/Hβcd-rGO TEM pictures and **(H)** SAED pattern images, and **(I-K)** $Cu_2S$/ODA-rGO TEM pictures and **(L)** SAED pattern images.

The quality and crystallinity of the composite materials were further investigated with the high-resolution Transmission Electron Microscopy (TEM) and Selected Area Electron Diffraction (SAED). **Figure. 5(A-C)** shows $Cu_2S$/rGO composite materials with $Cu_2S$ particles tightly integrated within the rGO layers, demonstrating efficient dispersion and interaction among the phases. In **Figure. 5(C),** the HRTEM image allows to define the lattice fringes with the interplanar d spacing of 0.076 nm and 0.069 nm that align with $Cu_2S$ crystal planes. In **Figure. 5(D)** in SAED image, a distinct diffraction pattern indicates the crystalline structure of $Cu_2S$ and the rGO, as well as their lattice co-orientation. In **Figure. 5(E-G)** $Cu_2S$ is embedded in Hβcd functionalized rGO, making the composite permeable and less aggregated. **Figure. 5(F)** shows well-dispersed $Cu_2S$ components within the stacked rGO structure, indicating significant interaction with Hβcd-rGO. The HRTEM image in the **Figure. 5(G)** reveals $Cu_2S$ structure with lattice spacings of 0.097 nm and 0.084 nm. The SAED pattern in **Figure. 5(H)** confirms the crystal parameters of the composite, also showing a strong diffraction ring, indicative of larger disorder. **Figure. 5(I-K)** displays $Cu_2S$ morphology with ODA-functionalized rGO. In **Figure. 5(J)** more homogeneous dispersion of $Cu_2S$ components in the ODA-rGO matrix can be traced. **Figure. 5(K)** verifies $Cu_2S$ crystal lattice fringes at 0.16 and 0.14 nm distances. **Figure. S5** depicts the TEM images of **(A-C)** rGO, **(D-F)** Hβcd-rGO, and **(G-I)** ODA-rGO. rGO is a layered material composed of individual sheets that are aggregated through van der Waals forces. Despite the functionalization with molecules such as Hβcd-rGO and ODA-rGO, the interlayer contacts are predominantly influenced by van der Waals forces, while supplementary hydrogen bonding or electrostatic attraction may contribute partially. The observed morphological features from the FESEM and TEM images show that hierarchical and defect-rich morphology increases the electrochemically active surface area and promotes rapid charge transfer, aligning with the observed rise in peak current and reduction in charge-transfer resistance during electrochemical analysis.

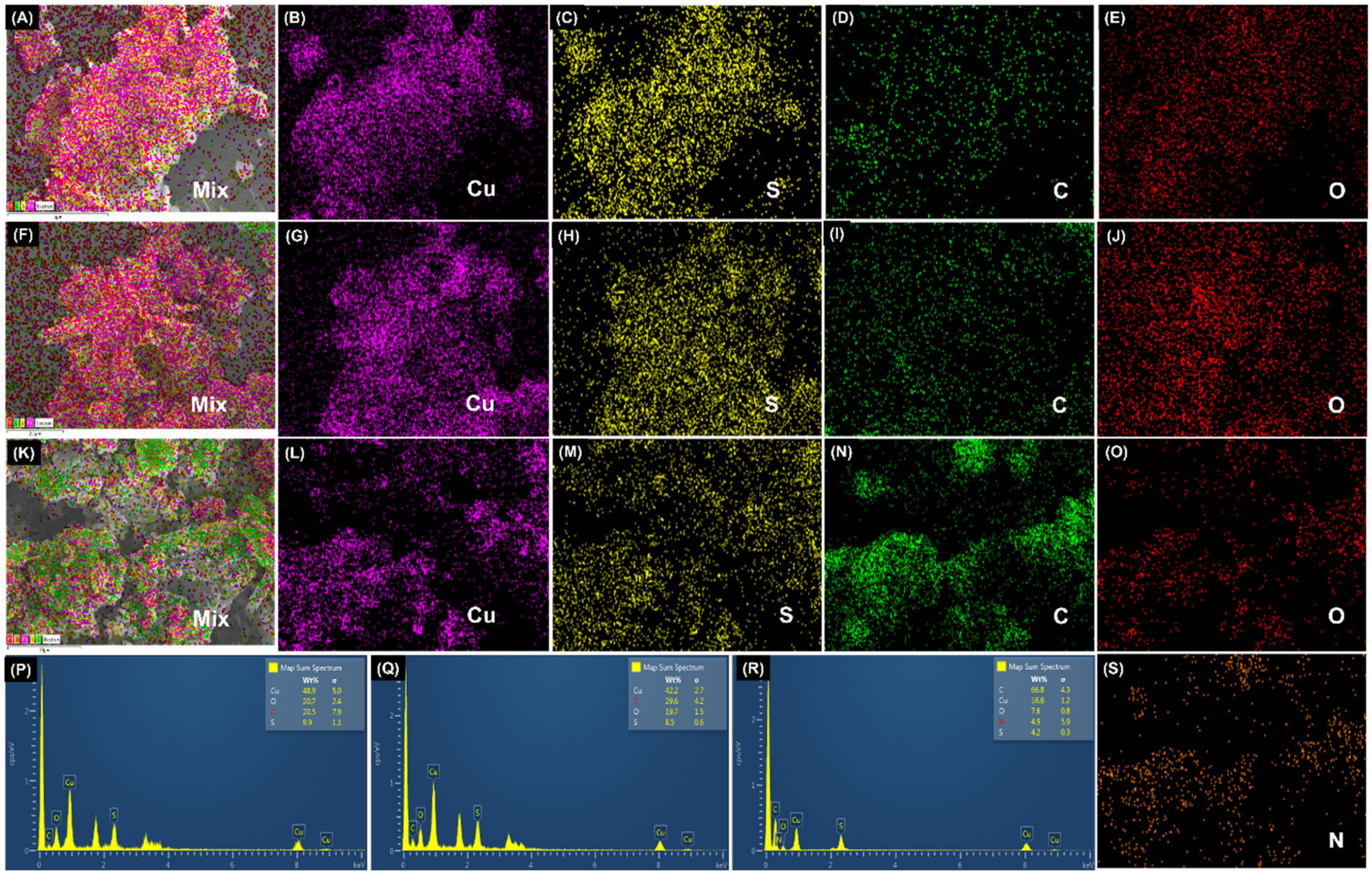

**Figure. 6** FESEM elemental mapping images of **(A, E)** $Cu_2S$/rGO, **(F-J)** $Cu_2S$/Hβcd-rGO, **(K-O and N)** $Cu_2S$/ODA-rGO, and **(P-R)** EDAX spectrum of $Cu_2S$/rGO, $Cu_2S$/Hβcd-rGO, and $Cu_2S$/ODA-rGO.

FESEM elemental mapping was used to evaluate the co-distribution of $Cu_2S$ and rGO phases in the composite materials, as well as to clarify the role of functionalization in the organic compounds. In **Figure. 6(A-E),** the mixed elemental mapping, combining copper, sulfur, carbon, and oxygen maps, shows the consistent dispersion of all of these components in the $Cu_2S$/rGO composite. Similar elemental mapping is shown for the $Cu_2S$/Hβcd-rGO composite in **Figure.**

**6(F-J)** and for $Cu_2S$/ODA-rGO in **Figure. 6(K-O, and S). Figure. 6(P-R)** represents the EDAX spectrum of $Cu_2S$/rGO, $Cu_2S$/Hβcd-rGO, and $Cu_2S$/ODA-rGO composites with atomic weight percentages.

The study regarding the quality of original $Cu_2S$ with FESEM analysis is presented in **Figure. S2(A-D), Figure S1(E-H),** and **Figure S4(A-J)** showing the elemental mapping and EDAX spectrum. **Figure. S3** shows the FESEM images for bare and functionalized rGO materials. **Figure. S4(K-M)** depicts the bare and functionalized rGO EDAX spectra with elemental content for bare rGO: C - 84.1 Wt % and O - 15.9 Wt %; for the Hβcd-rGO: C - 68.6 Wt % and O - 31.4 Wt %; and ODA-rGO: C - 61.0 Wt %, O - 27.7 Wt %, and N - 11.3 Wt %. The identification of an evenly distributed nitrogen signal confirms the effective functionalization of rGO with ODA. Hβcd-rGO exhibits elevated oxygen content, possibly due to the hydroxyl groups in the functionalization, similar to amine groups for ODA-rGO functionalization.

## 4. Electrochemical analysis:

### 4.1. Electrochemical resistance analysis of different electrodes

The electrochemical impedance spectroscopy (EIS) method was used to analyze the major electrochemical parameters of the GCE fabricated with the composite material, e.g., the charge transfer resistance. The value of the charge transfer resistance of the bare GCE and GCE modified with the prepared composite materials, reflects on the performance of the electrode – the higher the resistance the lower the electrocatalytic activity related to the diffusion of atoms that provoke the electron transfer at the electrode/electrolyte interface at the higher EIS frequencies. The EIS study was done with 5mM $[Fe(CN)_6]^{3-/-4}$ in 0.1 M KCl at a frequency range about 1 Hz to 1000 kHz and AC&DC potential values about 5 mV and 0.58 mV [54]. **Figure. 7(A)** presents the Nyquist plots obtained for bare and modified electrodes (an inset shows the Randles-Sevcik circuit diagram used for the data analysis). From the analysis the values of charge transfer resistance follow the sequence: bare GCE < $Cu_2S$/GCE < rGO/GCE < ODA-rGO/GCE < Hβcd-rGO/GCE < $Cu_2S$/rGOs/GCE < $Cu_2S$/ODA-rGOs/GCE < $Cu_2S$/Hβcd-rGO/GCE, with numerical resistance values from the data: 497 Ω < 410 Ω < 333 Ω < 318 Ω < 266.5 Ω < 150 Ω < 126 Ω < 61 Ω and respectively. The functionalized $Cu_2S$/Hβcd-rGO/GCE exhibits the lowest Rct value. The synergistic interaction between $Cu_2S$, which provides more active sites, and Hβcd-rGO, which helps to bind the analyte, results in improving the charge transfer kinetics and reducing the charge transfer resistance ($R_{ct}$) values. **Figure. 7(B)** displays the Rct histogram diagram from the EIS study.

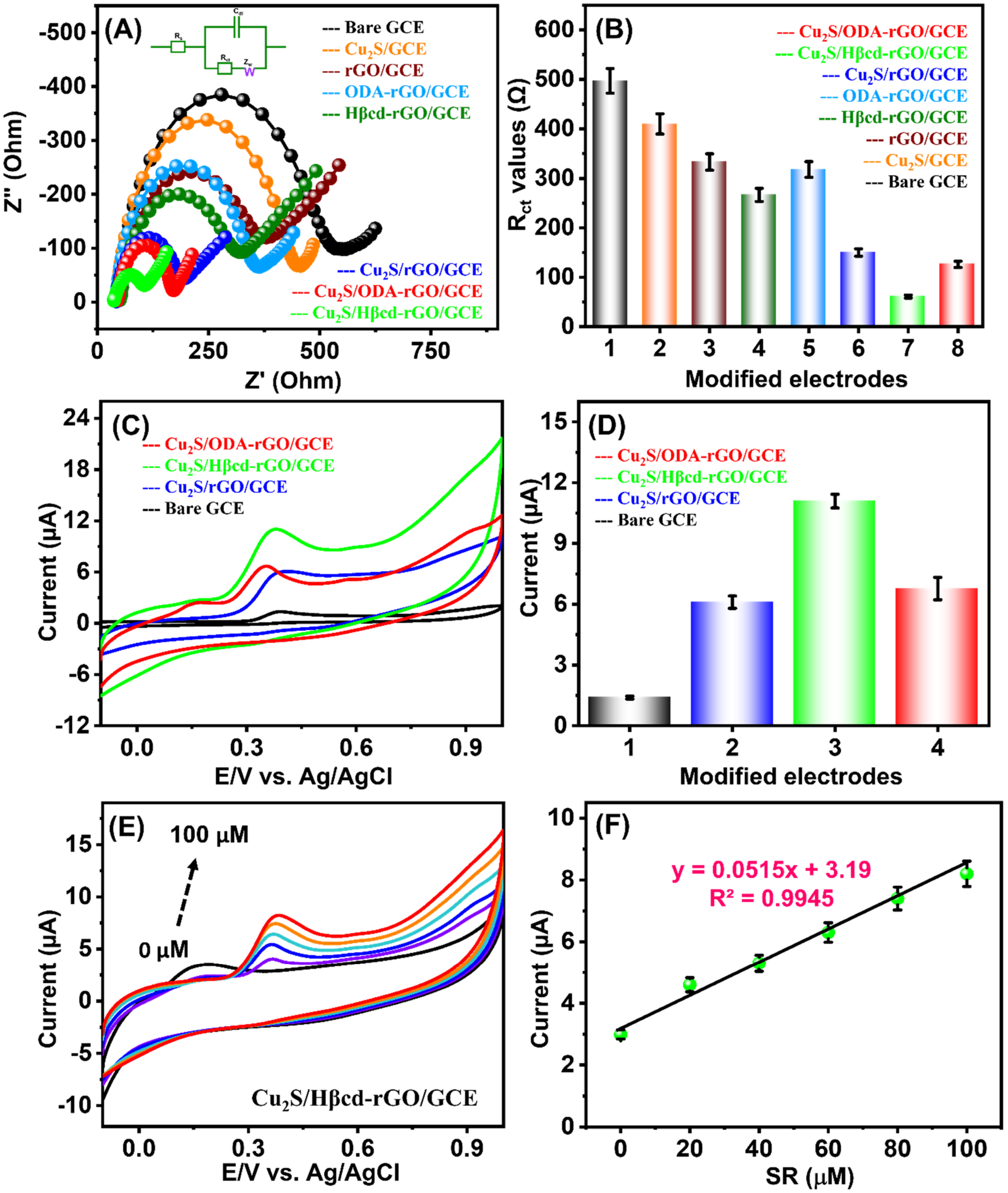

**Figure. 7(A)** EIS of bare GCE, $Cu_2S$/GCE, rGO/GCE, Hβcd-rGO/GCE, ODA-rGO/GCE, $Cu_2S$/rGO/GCE, $Cu_2S$/Hβcd-rGO/GCE, and $Cu_2S$/ODA-rGO/GCE in ferricyanide solution with a frequency range of 1 Hz to 1 MHz. **(B)** and its corresponding histogram plot **(C)** CV profiles of bare GCE, $Cu_2S$/rGO/GCE, $Cu_2S$/Hβcd-rGO/GCE, and $Cu_2S$/ODA-rGO/GCE in 0.1 M of PB solution (pH 7.0) with 50 μM of SR, **(D)** and its corresponding histogram chart **(E)** different concentrations (0 – 100 μM) of SR studied with $Cu_2S$/Hβcd-rGO/GCE in PB solution (pH – 7.0), and **(F)** linear fit of current *vs* SR concentration.

**4.2. Electrochemical detection of SR:**

The electrochemical sensing of SR was carried out with the bare GCE and modified $Cu_2S$/GCE, rGO/GCE, $Cu_2S$/rGO/GCE, Hβcd-rGO/GCE, ODA-rGO/GCE, $Cu_2S$/ODA-rGO/GCE, $Cu_2S$/Hβcd-rGO/GCE in 0.1M PBS (pH 7.0) at a scan rate of 50 mV/s upon addition of 150 μM SR. The obtained CV curves are illustrated in **Figure. 7(C)** with the histogram shown in **Figure. 7(D)** for bare GCE compared with the $Cu_2S$/rGO/GCE, $Cu_2S$/Hβcd-rGO/GCE, and $Cu_2S$/ODA-rGO/GCE. **Figure. S6(A, B)** displays the CV curves and their histogram plot compared with bare GCE and $Cu_2S$/GCE, rGO/GCE, Hβcd-rGO/GCE, and ODA-rGO/GCE. The bare GCE exhibits a broad peak with relatively lower peak current, which indicates the low performance towards the detection of SR at anodic current of 1.39 μA at 0.39 V. The GCE fabricated with $Cu_2S$ material shows an enhanced current response of 1.87 μA at a potential of 0.40 V compared to bare GCE. Subsequently, the GCE coated with modified rGO, Hβcd-rGO/GCE, and ODA-rGO/GCE shows an increased oxidation peak current due to the better electron transfer and the stronger interaction between the analyte and the electrode, which provides more active sites and higher conductivity. The oxidation peak current for rGO, Hβcd-rGO/GCE, and ODA-rGO/GCE was 2.91 μA, 5.1 μA, and 4.27 μA, respectively. The Hβcd-rGO/GCE provides a higher oxidation peak current compared to other modified electrodes. The GCE modified with $Cu_2S$/rGO shows improved current response – faster electron transfer process resulting in the higher anodic current of 6.1 μA with a potential shift of 0.40 V. The modified $Cu_2S$/ODA-rGOs/GCE displays a current of 6.77 μA, slightly higher than for $Cu_2S$/rGO. The GCE modified with $Cu_2S$/Hβcd-rGO exhibits a current response of 11.09 μA, the largest amongst other modified electrodes, and a slight potential shift at 0.36 V.

### 4.3. SR concentration test:

The electrochemical performance of modified $Cu_2S$/Hβcd-rGO was studied with the CV method in 0.1M PBS with the addition of SR at a scan rate of 50 mV/s. The CV graphs represent a good response upon the addition of SR from 0 to 100 μM. The increasing concentration of SR leads to a gradually increasing oxidation peak current without peak shift against the positive potential. **Figure. 7(E)** indicates excellent sensing performance of the modified electrode and the **Figure. 7(F)** displays the linear response of $y = 0.0548x + 2.64$ with the correlation coefficient $R^2 = 0.9945$.

### 4.4. Effect of the scan rate and pH of the electrolyte:

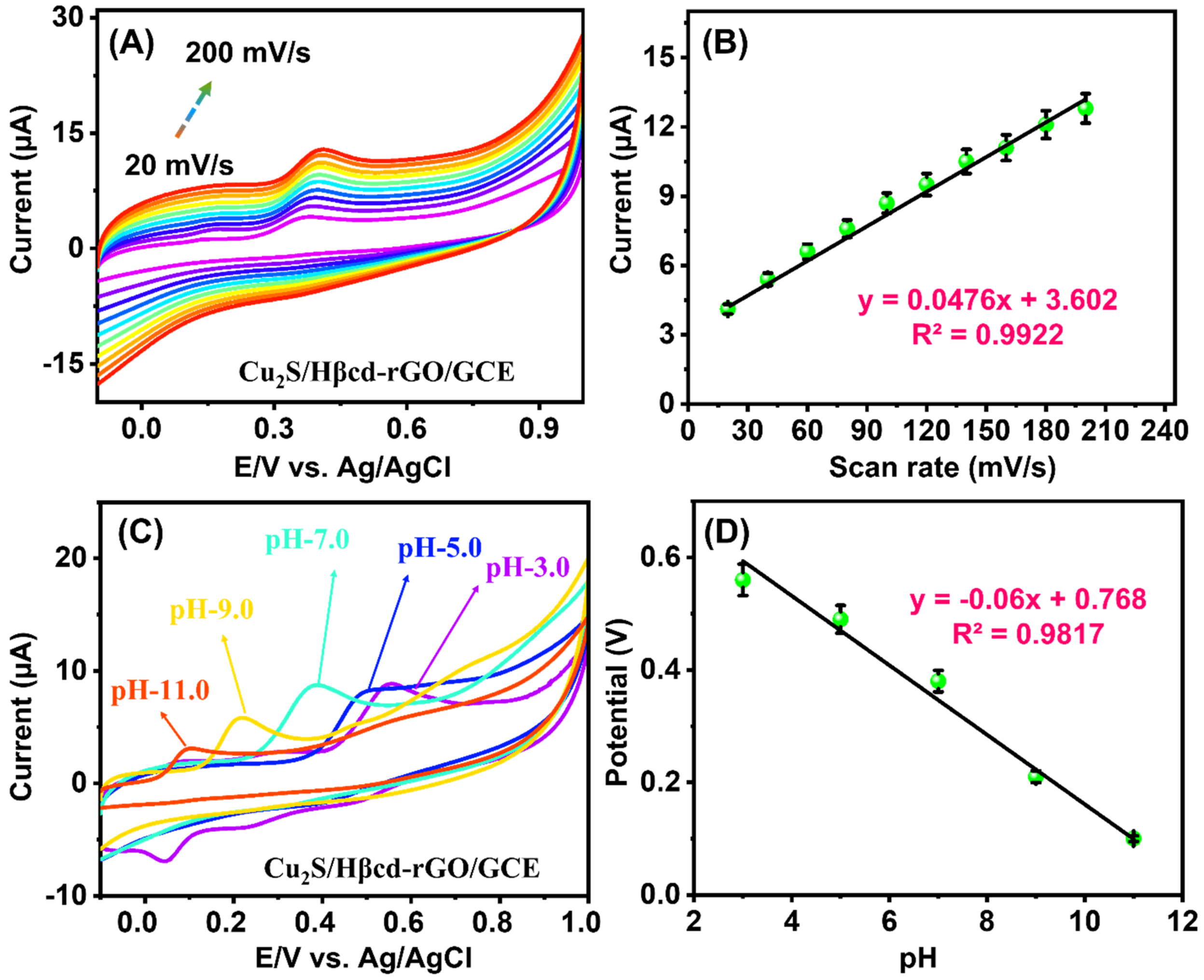

**Figure. 8(A)** CV profiles of $Cu_2S$/Hβcd-rGO/GCE with the addition of 50 μM of SR and a 0.1 M PB solution for different scan rates ranging from 20 mV/s to 200 mV/s**, (B)** a linear plot of the current values *vs* scan rates**, (C)** CV curves of $Cu_2S$/Hβcd-rGO/GCE at different pH levels (3.0 to 11.0) with a 75 μM of SR concentration conducted in a 0.1 M PB solution with a scan rate of 50 $mVs^{-1}$, **(D)** and the corresponding linear plot of the different pH levels *vs* peak potential values**.**

To determine the electron transfer over the modified $Cu_2S$/Hβcd-rGO/GCE, the CV technique was used with the 50 μM SR added in PBS (pH – 7) at different scan rates of 20 to 200 mV/s. Varying the scan rate, we observed an increase in the oxidation peak current, shifted towards the positive potential. The resultant CV curves vs. the scan rate are shown in **Figure. 8(A)** with the linear dependence of anodic peak current on the scan rates shown in **Figure. 8(B)** with the obtained linear equation as y = 0.0476 x + 3.602 and the correlation coefficient $R^2$ value 0.9922. The oxidation reaction process of SR at $Cu_2S$/Hβcd-rGO/GCE is largely controlled by an irreversible adsorption process. During the SR oxidation reaction, the number of electrons is nearly 2 (two-electron and two-proton oxidation process). The oxidation process of serotonin being converted into serotonin quinoneimine is illustrated in the **scheme. 2.** The oxidation peak of serotonin quinoneimine was identified at 0.36 V potential [42,62–64].

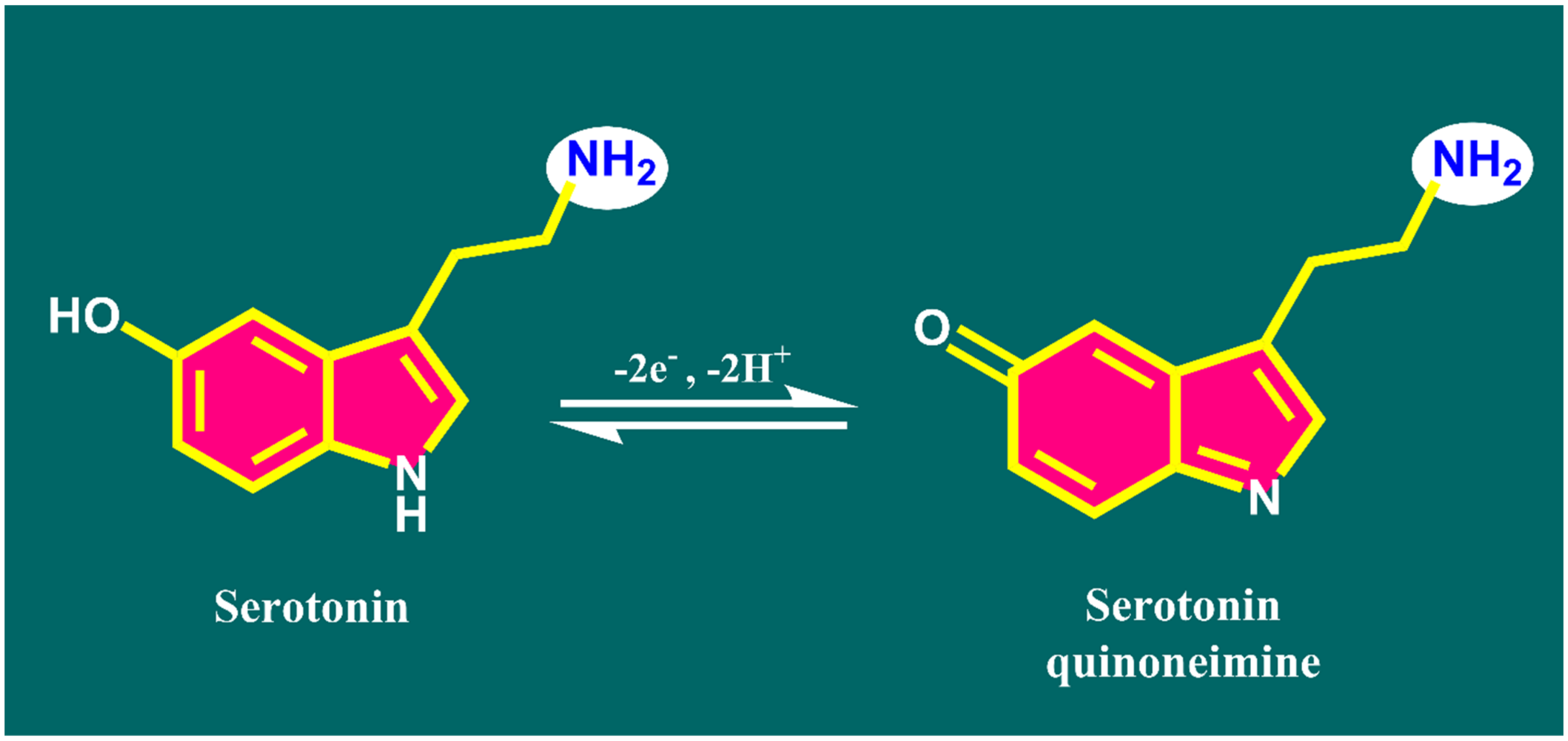

**Scheme .2** Oxidation detection mechanism of SR.

The determination of a suitable pH of the electrolyte solution employed for the electrochemical studies is essential. The study was done with 0.1 M PB solutions with various pH (3.0, 5.0, 7.0, 9.0, 11.0) upon addition of 75 µM SR at the constant scan rate of 50 mV/s. With increasing pH values, the oxidation peak potential shifts in the negative direction. **Figure. 8(C)** displays the CV curves and the **Figure. 8(D)** displays the linear fit for the oxidation peak potential vs. pH with the regression equation y = -0.060x + 0.768 (the correlation coefficient $R^2$ = 0.9817). The linearly increasing pH level in the range 3.0 to 7.0 results in the gradual increase of the anodic peak current, while in the pH range of 9.0 to 11.0, it results in a decrease of current response. Thus, pH 7.0 corresponds to the maximum oxidation peak current at the potential of 0.38 V.

### 4.5. DPV performance with the $Cu_2S$/Hβcd-rGO/GCE

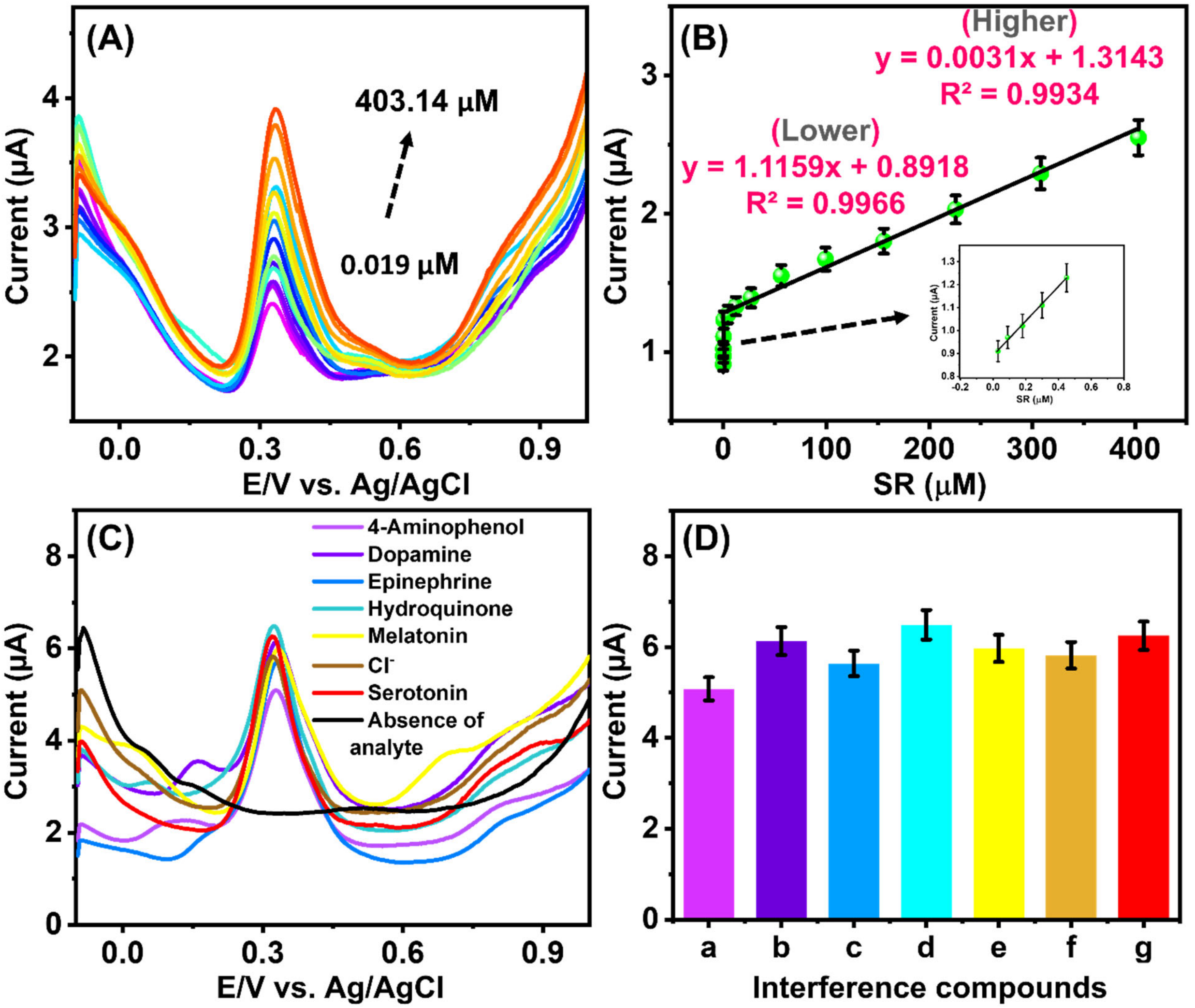

**Figure. 9(A)** DPV study at $Cu_2S$/Hβcd-rGO/GCE with varying concentrations of SR (0.019 to 0.299 and 4.28 to 403.14 μM) in 0.1 M PB solution, **(B)** the calibration plot showing the relationship between the concentration of SR and current, **(C)** selectivity investigation with $Cu_2S$/Hβcd-rGO/GCE using 75 μM SR injection together with the presence of several interferants, and **(D)** and its histogram plot of interferents vs current.

The differential pulse voltammetry (DPV) method was used to acquire a precise calibration of SR sensing, carried out at low SR concentrations. The DPV curves obtained upon addition of SR from lower concentration to higher concentration (in the range 0.019 to 0.299 μM and 4.28 to 403.14 μM) with the presence of $Cu_2S$/Hβcd-rGO/GCE are shown in **Figure. 9(A).** A linearly increasing oxidation peak current without potential shift indicates the enhanced sensing capacity of the modified electrode. **Figure. 9(B)** depicts the calibration plot for the oxidation peak current *vs* SR levels with the calibration fit of y = 1.11159x + 0.8919 and 0.0031x + 1.31148 (with $R^2$ = 0.9966 and 0.9934) for the addition of lower and higher SR concentrations. The low limit of detection (LOD) was calculated using the formula:

$$\text{Limit of detection } = 3\,\sigma/S$$

where σ is the standard deviation and S is the obtained slope value. The calculated LOD value was 0.0012 μM or 1.2 nM with the sensitivity about 15.9 μA $\mu M^{-1}$ $cm^{-2}$. The table comparing previous literature reported for SR detection with various modified electrodes is shown in **Table. 1**. **Table 1.** This table compares various modified electrode materials employed for monitoring SR with earlier published investigations.

| **Materials for different electrodes** | **Methods** | **Linear range (μM)** | **Detection of limit (μM)** | **Ref** |
|---|---|---|---|---|
| SPCE/MWCNT-S-Au | DPV | 1-12 | 1.0000 | [65] |
| GCE/AuNPs/AuNRTs-rGO | LSV | 3-100 | 0.3870 | [66] |
| GCE/MWCNT-NiO | SWV | 0.0598 – 62.8 | 0.1660 | [67] |
| Nafion-CNT/EC CFME | CV | 0.5 – 1.1 | 0.1400 | [68] |
| $Nb_2CT_x$/PCN/CCT | DPV | 1 - 100 | 0.0634 | [69] |
| AuNPs@PPyNPs | SWV | 0.1 - 15 | 0.0332 | [70] |
| RGO1/GCE | i-t | 1 - 100 | 0.0320 | [71] |
| rGO-$Ag_2Se$ | DPV | 0.1 - 15 | 0.0296 | [72] |
| SPCE-PPy-$Fe_3O_4$ NPsF | DPV | 0.007 – 0.1 | 0.0200 | [73] |
| FeC-AuNPs-MWCNT/SPCE | SWV | 0.05 - 20 | 0.0170 | [74] |
| $Ti_3C_2T_x$-rGO/GCE | DPV | 0.025 - 147 | 0.0100 | [35] |
| $Zn_2P_2O_7$/NbC/GCE | DPV | 0.019 – 563.68 | 0.0055 | [42] |
| $Cu_2S$/GCE | DPV | 0.029 – 607.6 | 0.0032 | [10] |
| $Cu_2S$/Hβcd-rGO/GCE | DPV | 0.019 to 0.299 and 4.28 to 403.14 µM | 0.0012 | **This work** |

**4.5. Anti-Interference and Reproducibility studies:**

To determine the selectivity of modified $Cu_2S$/Hβcd-rGO/GCE, an interference study was done with the presence of SR, along with other similar compounds in high quantities. The interference compounds, about 50 μM of each compound, such as aminophenol, dopamine, epinephrine, hydroquinone, melatonin, and chlorine, were used. Also, we have added additional metal ions(mercury) and phenolic compounds such as 4-nitrophenol, which are shown in **Figure S9.** The addition of SR at a 5-fold lower concentration than that of the other interfering compounds was added over the $Cu_2S$/Hβcd-rGO/GCE following the regular addition method. **Figure. 9(C)** illustrates the results with the histogram plot obtained for the same analysis shown in **Figure. 9(D).** While the peak current detection of the sensor based on the $Cu_2S$/Hβcd-rGO/GCE modified electrode showed responses at potentials ranging between 0 to 1.5 V for the interferants like aminophenol, hydroquinone, dopamine, and melatonin. These results are associated with the potential influence that occurred due to the addition of SR together. Even though the SR peak is observed with a higher current response, which deliberately shows the excellent selectivity analysis. The reproducibility study of modified $Cu_2S$/Hβcd-rGO/GCE was done with the CV method. As shown in the **Figure. S7(A, B)** five different measurements for the same amount of SR show insignificant variation of the anodic peak current, demonstrating good reproducibility of the $Cu_2S$/Hβcd-rGO/GCE for the SR sensing. The repeatability of the constructed electrode was assessed through five successive measurements, as depicted in Figure S8A. The tests were performed in a 0.1 M phosphate buffer solution having 50 μM SR, utilizing the $Cu_2S$/Hβcd-rGO/GC electrode under cyclic voltammetry parameters. The results demonstrated essentially similar anodic peak currents, signifying exceptional repeatability of the sensor's operation. The long-term stability of the $Cu_2S$/Hβcd-rGO/GCE was assessed during a 15-day duration, as illustrated in Figure S8B. The electrode was tested at first and subsequently stored in a refrigerator between readings. CV outcomes were documented at consistent intervals (0, 5, 10, and 15 days) in a 0.1 M phosphate buffer solution containing 50 μM SR. The results exhibited negligible fluctuation in current responsiveness, validating the excellent stability of the modified electrode throughout time.

**4.6. Real sample analysis:**

The real sample analysis was carried out in human blood serum 1 and human blood serum – 2 using the DPV technique. The human blood samples were acquired from Sigma Aldrich Taiwan. The purchased blood samples (50 μM) were diluted in 0.1 M PBS and taken for analysis. The prepared human blood samples were taken to detect SR. The performance was studied using the regular addition procedure. The resultant DPV responses are displayed in the **Figure. 10(A, B). Figure. 10(C, D)** shows the linear plot for the real sample investigation. The DPV curves demonstrate the efficiency of the $Cu_2S$/Hβcd-rGO/GCE for the real sample investigation. The **Table. 2** illustrates the real sample calculation of recovery and added range.

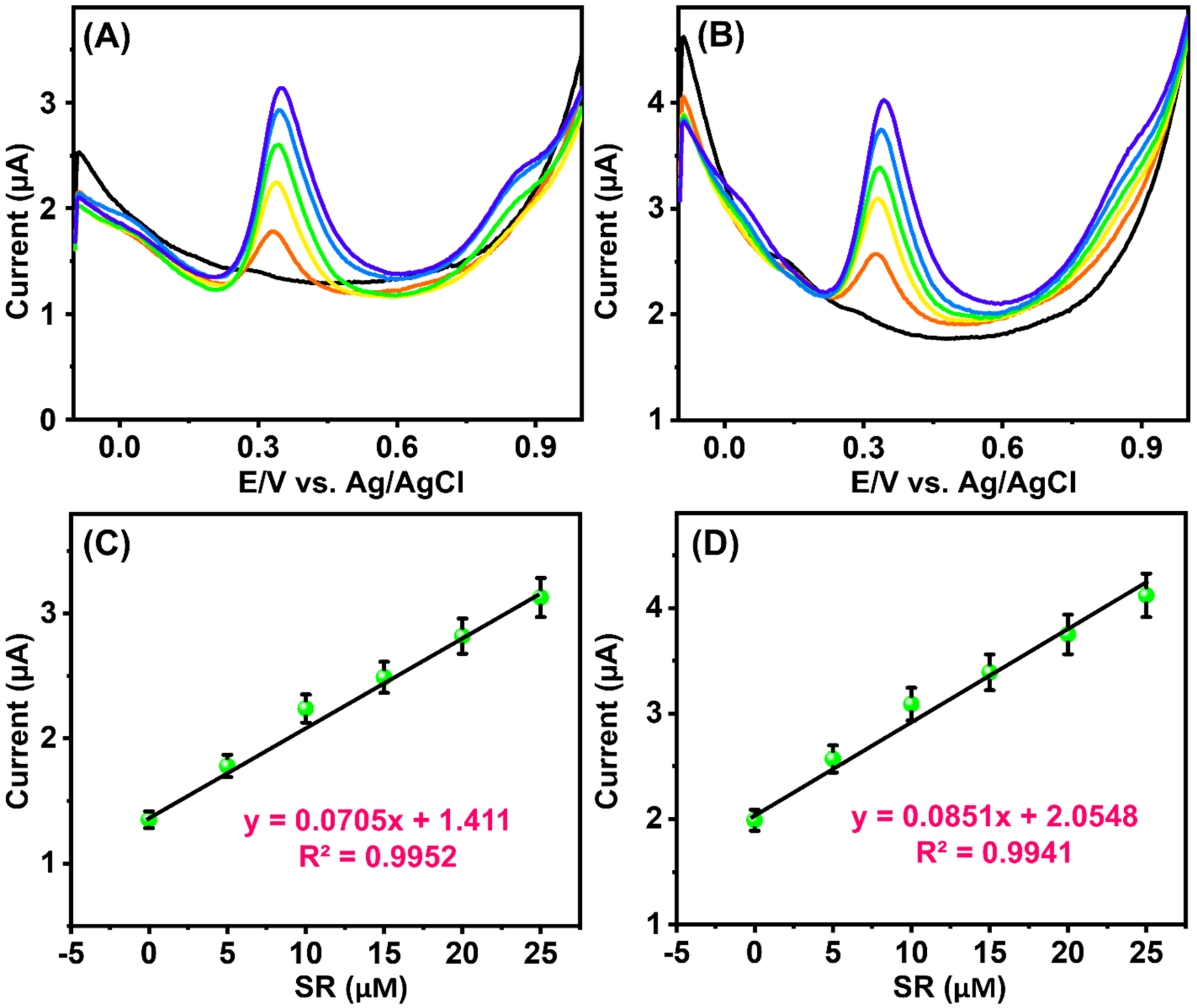

**Figure. 10** Real sample investigation of $Cu_2S$/Hβcd-rGO/GCE with the presence of SR, **(A)** human blood serum sample 1, **(B)** human blood serum-2, **(C, D),** and its linear plot of human blood serum-1 and 2.

**Table. 2.** Real sample analysis recovery percentage tabulation (n = 3).

| **Samples** | **Added (µM)** | **Detected (µM)** DPV | **Recovery (%)** (n=3) |
|---|---|---|---|
| | 0 | - | - |
| | 5 | 4.94 | 98.8 |
| **Human blood serum sample-1** | 10 | 9.87 | 98.7 |
| | 15 | 14.90 | 99.3 |
| | 20 | 19.91 | 99.5 |
| | 25 | 24.95 | 99.8 |
| **Human blood serum sample-2** | 0 | - | - |
| | 5 | 4.96 | 99.2 |
| | 10 | 9.90 | 99.0 |
| | 15 | 14.93 | 99.5 |
| | 20 | 19.94 | 99.7 |
| | 25 | 24.92 | 99.6 |

## 5. Conclusion

In summary, the copper sulfide-based nanocomposites were effectively synthesized by a cost-efficient hydrothermal method, including $Cu_2S$ with unmodified rGO, β-cyclodextrin-functionalized rGO (HβCD-rGO), and octadecylamine-modified rGO (ODA-rGO). Thorough physicochemical investigations (UV–Vis, FT-IR, XRD, Raman, XPS, FESEM, and HRTEM) validated the establishment of distinct layered structures with close interfacial interaction between $Cu_2S$ and the rGO matrices. Of the synthesized electrocatalysts, $Cu_2S$/HβCD-rGO demonstrated the superior electrochemical performance for SR detection, attaining a detection limit of 1.2 nM with remarkable sensitivity and selectivity. The improved performance arises from the synergistic interaction between $Cu_2S$ nanostructures and HβCD-functionalized rGO, which increases the availability of active sites, enhances charge transfer, and speeds redox kinetics. This distinctive host–guest interface promotes effective adsorption and oxidation of SR compounds while guaranteeing long-term stability and repeatability. The nanocomposite electrode demonstrated exceptional analytical recovery in actual biological samples (human serum 1 and 2), underscoring its significant potential for practical biosensing and diagnostic applications. This study presents a meticulously designed $Cu_2S$/HβCD-rGO hybrid electrode as a durable, sensitive, and scalable platform for enhanced electrochemical detection of sulfonated pharmaceuticals and associated analytes.

**Competing financial interests**: The authors declare no competing financial interests.

**Acknowledgments**

The project was supported by the NTUT-PSU seed grant. The authors would like to express gratitude for the financial support provided by the National Taipei University of Technology-Pennsylvania State University NTUT-PSU-113-03. The Precision Analysis and Materials Research Center of the National Taipei University of Technology supports the instrumentation.